\documentclass[preprint2]{aastex63}

\usepackage{hyperref}
\usepackage{cleveref}
\usepackage{graphicx}
\usepackage[utf8x]{inputenc}
\usepackage{auto-pst-pdf} 
\usepackage{url}
\usepackage{breakurl}

\graphicspath{{./}{figures/}}

\shorttitle{An old binary OC in the Galaxy}
\shortauthors{Denilso Camargo}

\begin{document}

\title{NGC1605a and b: an old binary open cluster in the Galaxy}

\correspondingauthor{Denilso Camargo}
\email{denilso.camargo@gmail.com}

\author{Denilso Camargo}
\affil{Col\'egio Militar de Porto Alegre, Minist\'{e}rio da Defesa\footnote{Civil staff} \\
Av. Jos\'{e} Bonif\'{a}cio 363, Porto Alegre, 90040-130, RS, Brazil}

\begin{abstract}
This work communicates the discovery of a binary open cluster within the Galaxy. NGC 1605 presents an unusual morphology with a sparse stellar distribution and a double core in close angular proximity.
The 2MASS and Gaia-EDR3 field-star decontaminated colour-magnitude diagrams (CMDs) show two distinct stellar populations located at the same heliocentric distance of $\sim2.6$ kpc suggesting that there are two clusters in the region, NGC 1605a and NGC 1605b, with ages of $2$ Gyr and  $600$ Myr, respectively.  Both Gaia parallax and PM distributions are compact and very similar indicating that they are open clusters (OCs) and share the same kinematics.  The large age difference, 1.4 Gyr, points to a formation by tidal capture during a close encounter and the close spatial proximity and similar kinematics suggest an ongoing merger event. 
There are some prominent tidal debris that appear to trace the clusters' orbits during the close encounter and, unexpectedly, some of them appear to be bound structures, which may suggest that additionaly to the evaporation the merging clusters are being broken apart into smaller structures by the combination of Galactic disk, Perseus arm, and mutual tidal interactions.
In this sense, the newly found binary cluster may be a key object on the observational validation of theoretical studies on binary cluster pairs formation by tidal capture as well as in the formation of massive clusters by merging, and tidal disruption of stellar systems.

\end{abstract}

\keywords{(Galaxy:) open clusters and associations: general, (Galaxy:) open clusters and associations: individual}

\rightline{\it Accepted for publication in the ApJ}

\section{Introduction} \label{sec:intro}

Most known open cluster (OC) pairs are young with some few intermediate-age exceptions and no one old survivor, which suggest that such a systems are short-lived \citep{Bhatia88, Bhatia91, Dieball02, Pietrzynski00, Bekki04, Minniti04, Priyatikanto16, Kovaleva20}. In this sense, the survival timescale for binary clusters appears to be in the range 10-100 Myr, with less than $17\%$ of the Galactic multiple systems surviving for more than 25 Myr \citep{Dieball02, Fuente09, Fuente10}.

\begin{figure*}
\centering
\begin{minipage}[b]{1.0\linewidth}
\begin{minipage}[b]{0.3\linewidth}
\resizebox{\hsize}{!}{\includegraphics{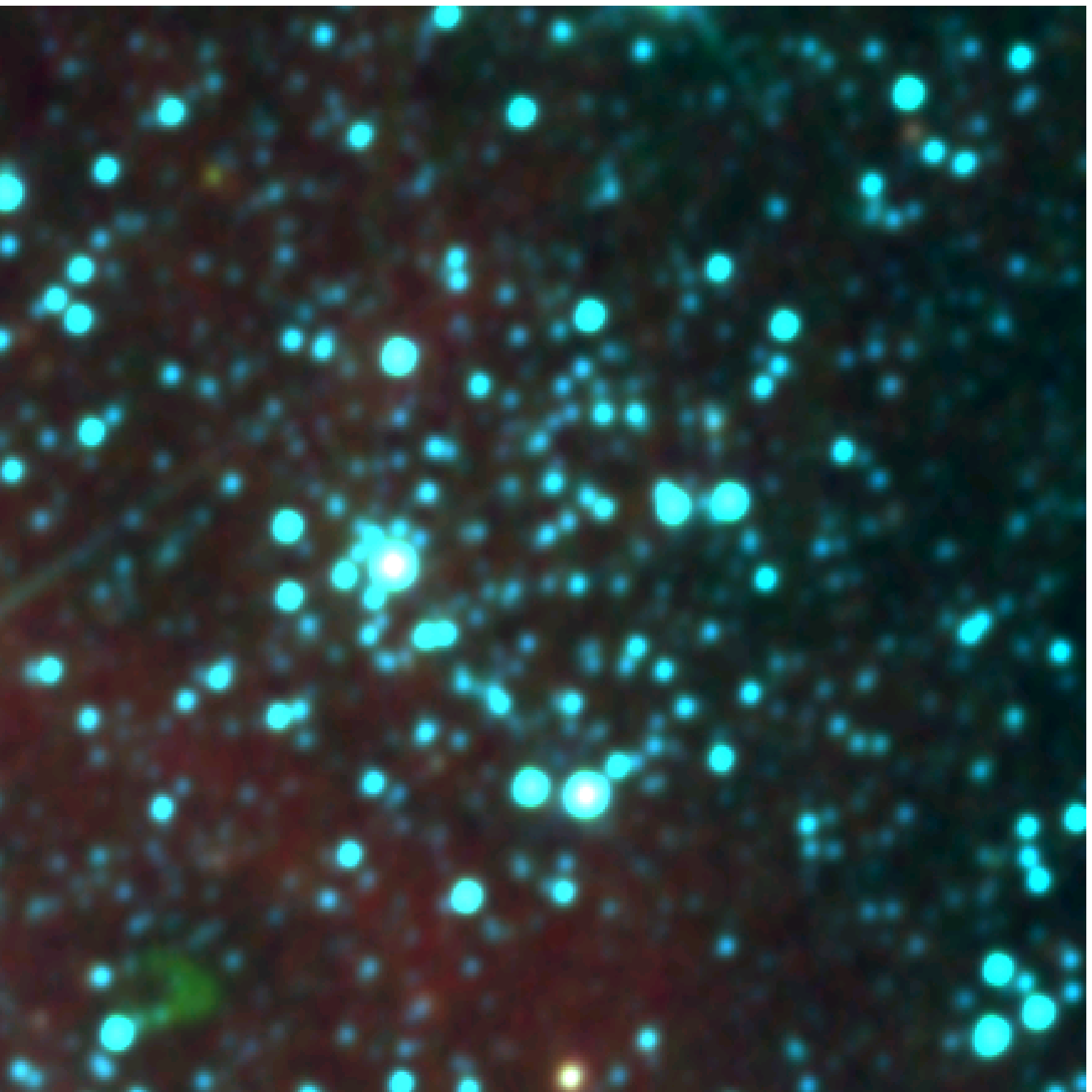}}
\put(-60.0,107.0){\makebox(0.0,0.0)[5]{\fontsize{14}{14}\selectfont \color{red} a}}
\put(-108.0,90.0){\makebox(0.0,0.0)[5]{\fontsize{14}{14}\selectfont \color{red} b}}
\end{minipage}\hfill
%\vspace{-0.2cm}
\begin{minipage}[b]{0.31\linewidth}
\includegraphics[width=\textwidth]{NGC1605circb.eps}
\end{minipage}\hfill
\begin{minipage}[b]{0.37\linewidth}
\includegraphics[width=\textwidth]{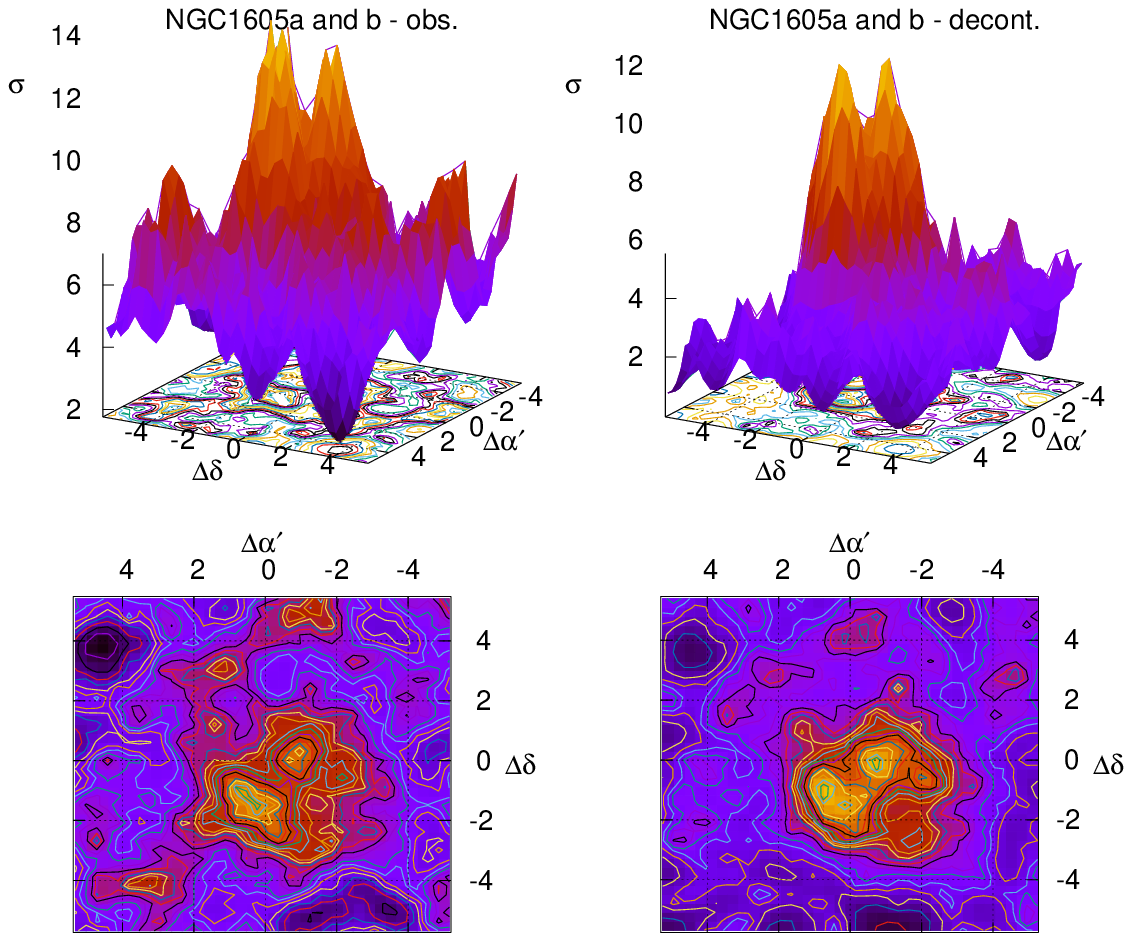}
\end{minipage}\hfill
\caption{\textit{Left panel}: WISE multicolour image ($10'\times10'$) of the cluster pair region. \textit{Middle panel}:  schematic distribution of star in the 2MASS observed (gray) and decontaminated (black) photometry. \textit{Right panels}: Gaia-EDR3 photometry. \textit{Top} - stellar surface-density $\sigma (stars\, {arcmin}^{-2})$ for the binary cluster pair region. \textit{Bottom} - the corresponding isopleth surfaces. \textit{Left} - observed photometry. \textit{Right} - decontaminated photometry. $\Delta{\alpha}'=\Delta(\alpha\,cos(\delta))$. The isopleth surfaces show evidence of tidal streams around the cluster pair.}
\label{f1}
\end{minipage}\hfill
\end{figure*}

Binary clusters may form after the gravitational collapse of multiple massive and dense gas clumps inside giant molecular clouds, with collapsing clumps forming clusters of similar ages and very close to each other \citep{Piatti10, Camargo11, Camargo15b, Arnold17, Mora19, Darma21}. However, like the individual embedded clusters multiples also suffer with the \text{infante mortality} \citep{Lada03} and, thus, most end up completely unbound or merge within a few tens of Myr. On the other hand, occasionally, a cluster pair may be formed by tidal capture during a close encounter between two clusters tipically with different ages \citep{van96, Oliveira00, Fuente09}.

During a close encounter the interacting clusters loss angular momentum and orbital energy, carried away by escaping stars and tidal interactions and, thus, they may become gravitationally bound together to form a binary cluster pair that eventually merge to compose a single cluster with two distinct stellar populations. The color-magnitude diagram (CMD) for the merger product is a composite of the two populations \citep{Leon99}. Alternatively, for high relative velocities, they can just perform a fly-by or end up completely disrupted. 
Also close encounters increase the internal energy of each cluster leading to radial expansion and enhancing the fraction of escaping stars. Such an event produces a rotating merger remnant, which also increase the mass-loss rate \citep{Einsel99, Fuente10, Priyatikanto16}.{\bf
It is expected that the escaping low mass stars form tidal streams tracing the clusters path.} 

While tidal streams from disrupted massive globular clusters and dwarf galaxies are relatively common in the galactic halo, tidal streams from disrupted lower mass open clusters in the galactic disk have been more difficult to detect \citep{Meingast19, Roser19, Barba19, Gagne21, Jerabkova21, Kamdar21}. 

In the following, is shown that open cluster NGC 1605 is in fact two merging OCs (hereafter, NGC 1605a and NGC 1605b) with evidence of tidal streams.
Four prominent overdensities were identified within the tidal streams using the Gaia photometry \citep{Gaia21}.
The paper is organized as follows. In Section 2, the methods and tools  employed in the clusters analyses are described. The Section 3, presents the fundamental and structural parameters derived. In Section 4, the results are discussed. Finally, the Section 5 provides the concluding remarks.

\begin{figure}
\resizebox{\hsize}{!}{\includegraphics{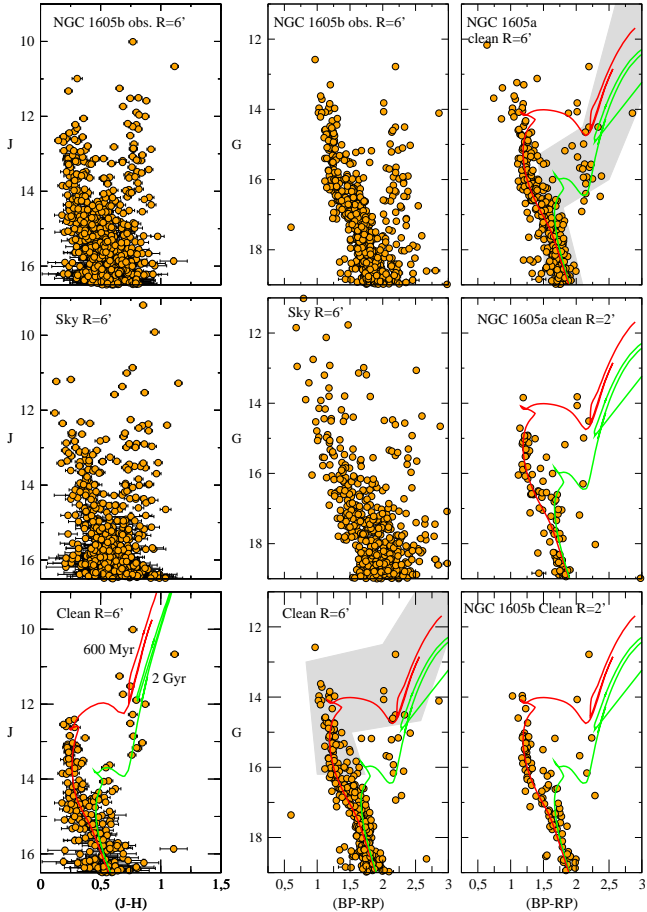}}
\caption{\textit{First column panels}: 2MASS $J\times(J-H)$ CMDs ($R=6'$) centered in NGC 1605b.  \textit{Top panel}: observed CMDs. \textit{Middle panel}: equal-area comparison field used only to illustrate the local field.  \textit{Bottom panel}: field-star decontaminated CMD fitted with Padova isochrones of 600 Myr and 2 Gyr. \textit{Second column  panels}: the same as the first column for the Gaia-EDR3 $G{\times}(BP-RP)$ CMDs. \textit{Third column of panels}: Gaia-EDR3 $G{\times}(BP-RP)$ CMDs. \textit{Top panel} - decontaminated CMD centered in NGC1605a ($R=6'$). \textit{Middle panel} - the same as the top panel for $R=2'$. \textit{Bottom panel} - decontaminated CMD of NGC1605b ($R=2'$). Note that CMDs with extractions of $R=6'$ show both OCs while those with $R=2'$ present the OCs individually.
Colour-magnitude filters used to isolate cluster stars are shown as shaded regions.}
\label{f2}
\end{figure}

\section{Methods}
\label{sec:2}

The cluster pair analysis is based on the 2MASS and Gaia-EDR3 photometry and WISE image.
The 2MASS and Gaia color magnitude diagrams (CMDs), built with photometry extracted in circular concentric regions centered in the optimised coordinates, provide the clusters basic parameters.

\subsection{Field-stars decontamination}
\label{sec:2a}
To uncover the cluster intrinsic CMDs it is employed a field-stars decontamination procedure \citep[][]{Bonatto07, Camargo15a, Camargo15b} previously applied in the 2MASS \citep{Camargo12, Camargo18}, WISE \citep{Camargo16}, and Gaia \citep{Minniti21b} photometry. 
The FS decontamination tool divides the CMDs of the cluster region and of a comparison field into a 3D grid (a magnitude and two colors), computing the density of stars with compatible magnitude and colors in the correspondent cells, and then, statisticaly remove from the cluster CMD the expected number of field stars in each cell. Large extraction areas are essential for a consistent field-star decontamination. 
For NGC1605a and b a wide external ring ($50'\leq{R}\leq70'$) centered in the clusters coordinates were adopted as comparison field.

\subsection{Fundamental parameters}
\label{sec:2b}
The basic parameters are derived by fitting PARSEC isochrones \citep{Bressan12} on the FS decontaminated CMDs. 

The fits are made by eye and consist in the application of shifts in magnitude and color in the isochrone from the zero distance modulus and reddening until a satisfactory solution is reached. The best fits are superimposed on decontaminated CMDs.
The clusters' parameters were derived by using both 2MASS and Gaia photometry independently.

\subsection{Structural parameters}
\label{sec:2c}

Structural parameters have been derived by means of the radial density profiles (RDPs), defined as the projected number of stars per area around the cluster center.
The RDPs of the OCs are built by applying color-magnitude filters to the Gaia EDR3 observed photometry. CM filters are used to exclude probable field stars, however, those with colors similar to that of the cluster's stellar sequences remain. The CM filter enhances the contrast of the RDP relative to the background level \citep{Camargo11, Camargo18, Camargo19}.
In order to guarantee RDPs with enough spatial resolution and moderate $1\sigma$ Poisson errors they were built by counting stars in concentric rings - the width of the rings grow with distance from the center. The structural parameters are derived by fitting a King function \citep{King62}. The King profile describes an isothermal sphere like bound stellar systems that reached a state of equilibrium.

The profile is expressed as $\sigma(R)=\sigma_{bg}+\frac{\sigma_{0K}}{[1 + (\frac{R}{R_c})^2]}$, where $\sigma_{bg}$ is the stellar background surface density, $\sigma_{0K}$ is the central density relative to the background level and $R_c$ is the core radius. The $\sigma_{bg}$ is measured in the comparison field and kept fixed to minimize degrees of freedom. 
 The cluster radius ($R_{RDP}$) and uncertainty can be estimated by considering the fluctuations of the RDPs with respect to the residual background and  corresponds to the distance from the cluster center where RDP and comparison field become statistically indistinguishable.

\section{Results}
\label{sec:3}

NGC 1605 was previously considered as a 5 Gyr old OC located at a heliocentric distance of 1.1 kpc \citep{Sujatha03}. However, the Gaia EDR3 and 2MASS decontaminated  CMDs of the cluster projected region show two distinct stellar populations suggesting that NGC 1605 may be a composite of two clusters. 
The WISE image (Fig.~\ref{f1} - left panel) shows a sparse stellar distribution with more than a visual overdensity, which can be more easely identified in the observed and decontaminated photometry from 2MASS and Gaia-EDR3 (Fig.~\ref{f1} - middle and right panels). 

Since that the RDP of the original cluster present a dip at the center \citep{Barkhatova61} new central coordinates were searched. Two main overdensities were found and some other less prominent, which also point to a cluster pair with the minors stellar density peaks beeing, possibly, a consequence of dynamical interactions. The optimised coordinates of NGC1605a and NGC1605b are shown in Table~\ref{tab1}. A cluster (FSR 694) was found by \citet{Froebrich07} in the ​​NGC1605 projected area, but as the coordinates do not match the classical name was kept.

The best-fitting PARSEC isochrones in the 2MASS decontaminated  $J\times(J-H)$ and Gaia EDR3 $G\times(BP-RP)$ CMDs independently provide an age of $2\pm0.2$ Gyr for NGC 1605a and $600\pm100$ Myr for NGC 1605b (Fig.~\ref{f2}). 
A heliocentric distance of $d_{\odot}=2.6\pm0.4$ kpc was derived for both, which agree with the value obtained by \citet{Fang70}. As the clusters centers have a projected separation of only $\sim1.8$ pc, the stellar content of both objects seems to be mixed within the cluster pair area even for the central region, as shown by their CMDs (Fig.~\ref{f2}).

\begin{figure}
\resizebox{\hsize}{!}{\includegraphics{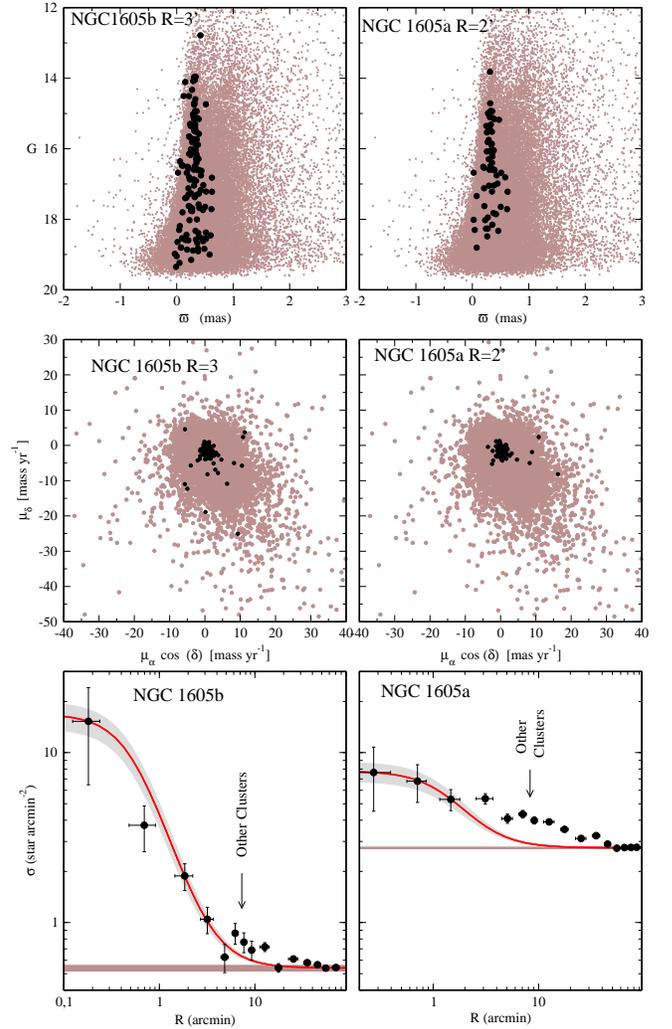}}
\caption{\textit{Top panels}: parallax for the Gaia EDR3 field-stars decontaminated photometry in the central region of each cluster and for field stars within a large area ($50'\leq{R}\leq70'$) in the cluster pair's neighborhood centered in NGC1605b.
  \textit{Middle panel}: Gaia EDR3 PM distribution. The black circles are the stars in the central region of each cluster, while the brown circles represent the stars in the same surrounding field  of the top panels. 
  \textit{Bottom panel}: radial density profile for NGC 1605a and NGC 1605b built with EDR3 CM-filtered photometry. Brown horizontal shaded region: stellar background level measured in the comparison field. Gray region: $1\sigma$ King fit uncertainty.}
\label{f3}
\end{figure}

The Gaia FS decontaminated photometry for the clusters' central regions present compact and very similar parallax distributions (Fig.~\ref{f3}). In addition, the Gaia-EDR3 PM distributions for the clusters' inner regions also are very similar and consistent with that of a typical open cluster. Fig.~\ref{f3} presents more compact PM distributions for stars in the clusters' area than in the comparison field, suggesting cohesive structures.

The Fig.~\ref{f3} also shows the RDPs for the two OCs. 
Structural parameters are displayed in Table~\ref{tab1}.
Since the stellar contents of the merging clusters appear to be fully mixed, as shown by the CMDs (Fig.~\ref{f2}), their RDPs present irregularities mostly due to the presence of the neighbor in the profile of each one of them. The CM filters cannot completely isolate the two stellar populations, as they overlap in the CMD, but for NGC 1605b it works better decreasing significantly the Galactic stellar field contribution to the cluster RDP. It seems that the field stars level in the NGC 1605b RDP represents the Galactic field in the pair neighborhood. The RDPs also show that the clusters stellar contents are spread in a large area.

\begin{figure}
\centering
\begin{minipage}[b]{1.0\linewidth}
\begin{minipage}[b]{1.0\linewidth}
\resizebox{\hsize}{!}{\includegraphics{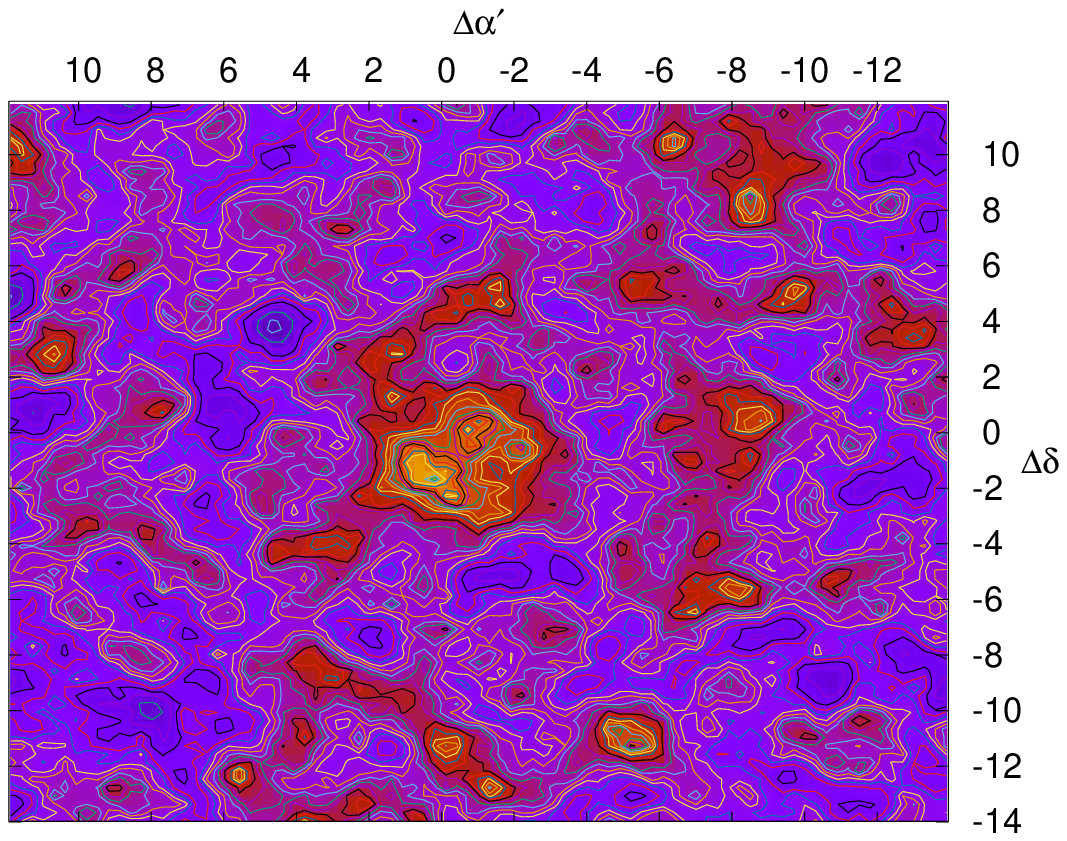}}
\put(-62.0,130.0){\makebox(0.0,0.0)[5]{\fontsize{14}{14}\selectfont \color{green} 1}}
\put(-62.0,88.0){\makebox(0.0,0.0)[5]{\fontsize{14}{14}\selectfont \color{green} 2}}
\put(-65.0,55.0){\makebox(0.0,0.0)[5]{\fontsize{14}{14}\selectfont \color{green} 3}}
\put(-130.0,35.0){\makebox(0.0,0.0)[5]{\fontsize{14}{14}\selectfont \color{green} 4}}
\end{minipage}\hfill
%\vspace{-1.5cm}
\begin{minipage}[b]{0.99\linewidth}
\includegraphics[width=\textwidth]{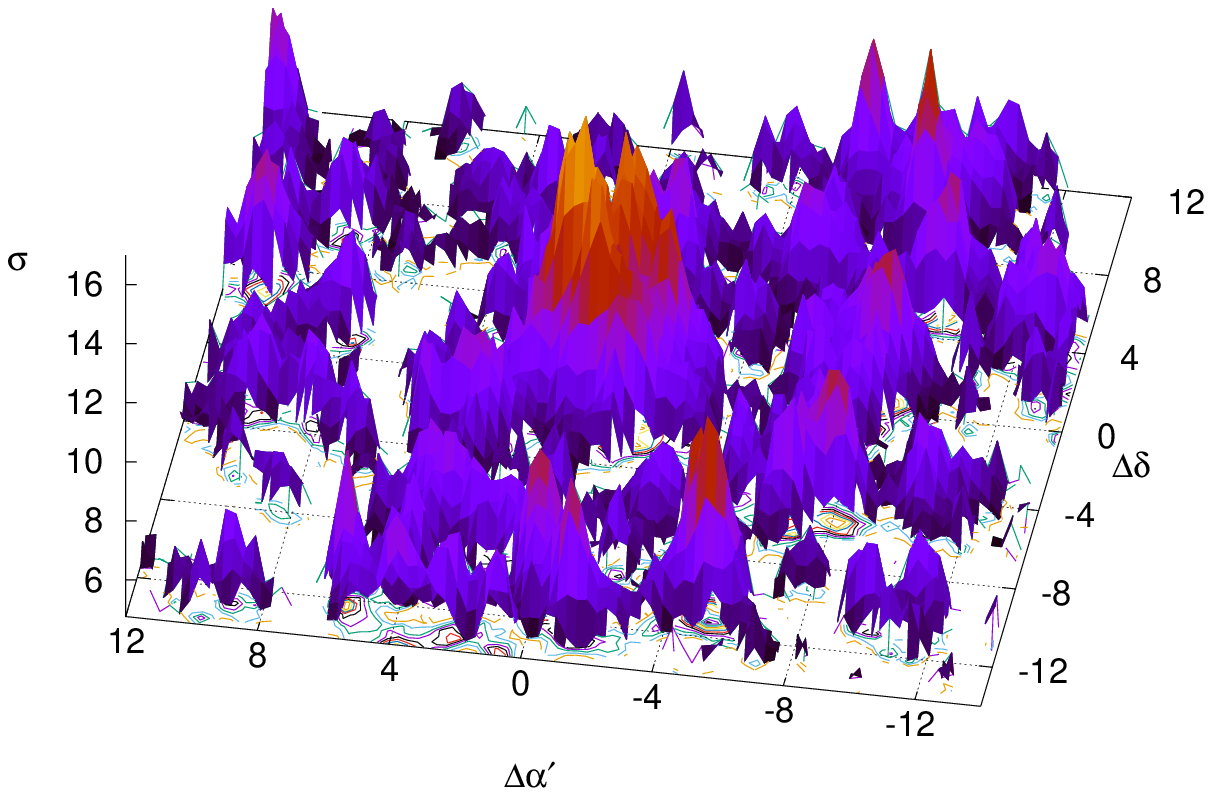}
\end{minipage}\hfill
\begin{minipage}[b]{0.96\linewidth}
\includegraphics[width=\textwidth]{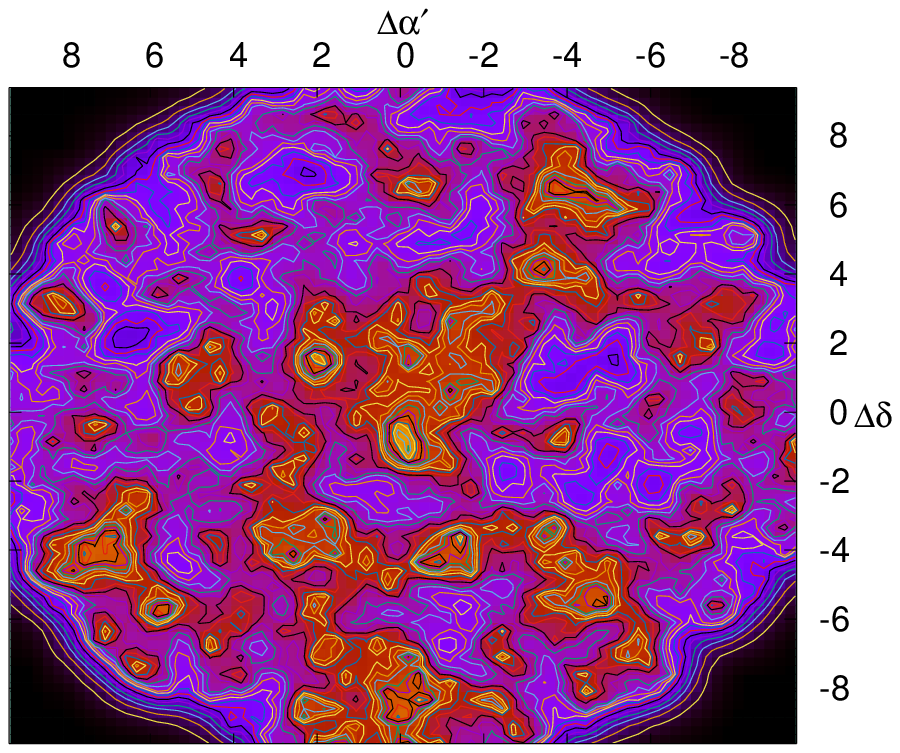}
\put(-100.0,90.0){\makebox(0.0,0.0)[5]{\fontsize{14}{14}\selectfont \color{green} 1}}
\put(-110.0,37.0){\makebox(0.0,0.0)[5]{\fontsize{14}{14}\selectfont \color{green} 2}}
\put(-169.0,143.0){\makebox(0.0,0.0)[5]{\fontsize{4}{4}\selectfont \color{green}$\bullet$}}
\put(-172.0,154.0){\makebox(0.0,0.0)[5]{\fontsize{4}{4}\selectfont \color{green}$\bullet$}}
\put(-163.0,162.0){\makebox(0.0,0.0)[5]{\fontsize{4}{4}\selectfont \color{green}$\bullet$}}
\put(-150.0,166.0){\makebox(0.0,0.0)[5]{\fontsize{4}{4}\selectfont \color{green}$\bullet$}}
\put(-133.0,162.0){\makebox(0.0,0.0)[5]{\fontsize{4}{4}\selectfont \color{green}$\bullet$}}
\put(-45.0,105.0){\makebox(0.0,0.0)[5]{\fontsize{4}{4}\selectfont \color{green}$\bullet$}}
\put(-60.0,117.0){\makebox(0.0,0.0)[5]{\fontsize{4}{4}\selectfont \color{green}$\bullet$}}
\put(-40.0,85.0){\makebox(0.0,0.0)[5]{\fontsize{4}{4}\selectfont \color{green}$\bullet$}}
\put(-64.0,55.0){\makebox(0.0,0.0)[5]{\fontsize{4}{4}\selectfont \color{green}$\bullet$}}
\put(-70.0,44.0){\makebox(0.0,0.0)[5]{\fontsize{4}{4}\selectfont \color{green}$\bullet$}}
\put(-76.0,31.0){\makebox(0.0,0.0)[5]{\fontsize{4}{4}\selectfont \color{green}$\bullet$}}
\put(-90.0,24.0){\makebox(0.0,0.0)[5]{\fontsize{4}{4}\selectfont \color{green}$\bullet$}}
\put(-100.0,65.0){\makebox(0.0,0.0)[5]{\fontsize{4}{4}\selectfont \color{green}$\bullet$}}
\put(-50.0,67.0){\makebox(0.0,0.0)[5]{\fontsize{4}{4}\selectfont \color{green}$\bullet$}}
\put(-143.0,26.0){\makebox(0.0,0.0)[5]{\fontsize{4}{4}\selectfont \color{green}$\bullet$}}
\put(-140.0,50.0){\makebox(0.0,0.0)[5]{\fontsize{4}{4}\selectfont \color{green}$\bullet$}}
\end{minipage}\hfill
\caption{\textit{Top panel:} stellar density map for the Gaia-EDR3 observed photometry centered in the binary cluster. \textit{Middle panel:} stellar surface-density $\sigma (stars\, {arcmin}^{-2})$ for the binary cluster pair region.
\textit{Bottom panel}: stellar density map showing details of the overdensities 1 and 2 with the secondary looping path (green dots). These features show clearly the orbital path of the merging clusters and the tidal debris within these tidal streams. $\Delta{\alpha}'=\Delta(\alpha\,cos(\delta))$.}
\label{f5}
\end{minipage}\hfill
\end{figure}

\begin{figure*}
\centering
\begin{minipage}[b]{0.9\linewidth}
\resizebox{\hsize}{!}{\includegraphics{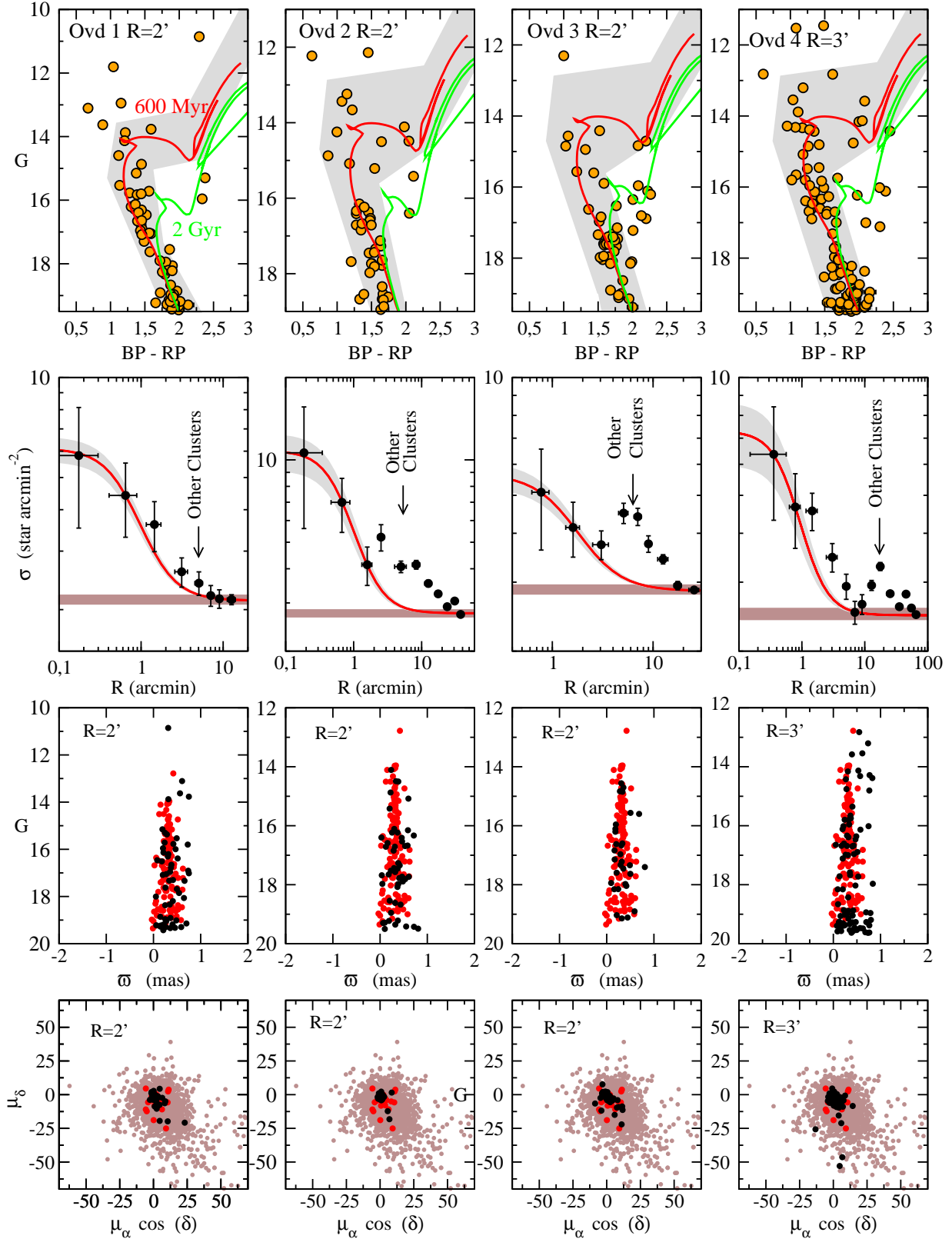}}
\end{minipage}\hfill
\caption{
\textit{First row panels}: Gaia-EDR3 decontaminated CMDs fitted with the same Padova isochrones of Fig.~\ref{f2}. CM filters used to isolate clusters stars are shown as shaded regions.
\textit{Second row panels}: RDPs built after applying CM filters that select the clusters sequences fitted by the isochrones.
\textit{Third row panels}: Gaia-EDR3 decontaminated parallax diagram. The black circles are the stars in the central region of each cluster and the red circles are the NGC1605b parallax distribution shown in Fig.~\ref{f3}.
\textit{Fourth row panels}: Gaia-EDR3 PM distributions. The black circles are the stars in the central region of each cluster, while the brown circles represent the stars in the surrounding field ($50'\leq{R}\leq70'$) and the red circles are the NGC1605b PM distribution shown in Fig.~\ref{f3}.
}
\label{f6}
\end{figure*}

Additionally, the Fig.~\ref{f1} provides the stellar surface-density and the corresponding isopleth surfaces for the Gaia EDR3 observed and decontaminated photometry, which point to a binary cluster and support an ongoing merger event. 
This figure shows clearly the presence of two prominent stellar clumps that can still be identified individually, but in an advanced merger stage. 
Since there are evidences of tidal debris in the binary cluster pair vicinity (Fig.~\ref{f1}), the isopleth surface and stellar surface-density for a large region are shown in Fig.~\ref{f5}. 
It is evident the tidal debris distribution along the tidal streams that trace the probable orbital paths of the two clusters converging in the present position of such objects. 
The analysis of the most prominent overdensities inside the tidal streams shows that at least some of them are populated by stars consistent with those that make up the binary cluster pair, mainly NGC1605b. 
The Fig.~\ref{f6} provides the Gaia EDR3 $G\times(BP-RP)$ CMDs, the RDPs, the decontaminated parallax diagrams, and PM distribution for four tidal debris overdensities. Since the Gaia parallax uncertainties may be underestimated by up to $\sim40\%$ \citep{Arenou18}  the decontaminated parallax diagrams were built using stars with uncertainties lower than $0.5\,mas$ and a more restrective decontamination procedure.
The best isochrone solutions for NGC1605a and NGC1605b fit very well the two populations inside these overdensities indicating that the tidal debris are possibly OCs located at the same heliocentric distance ($\sim2.6$ kpc) as the merging clusters, which may suggest a common origin. 
The age derived for NGC1605b (600 Myr) was adopted for the overdensities, but a significant fraction of their stars are consistent with those within NGC1605a indicating that the exchange of stars probably also took place earlier.

The RDPs of these overdensities follow a King law, which suggest that they may be  bound structures. Table~\ref{tab1} shows the structural parameters derived for these objects. 
The compact PM distributions also point to gravitationally bound structures (Fig.~\ref{f6}).

The analysis of the newly found overdensities is robust enough to classify them as OCs.
Following the star cluster catalog \citep{Camargo15a, Camargo16, Camargo18, Camargo19} these tidal debris OCs are also named Camargo 1010, 1011, 1012, and 1013.

\section{Discussions}
\label{sec:4}

Given that NGC1605a and NGC1605b are located at the same heliocentric distance and in close projected proximity, it seems that they form an old binary OC pair in the Galaxy. These objects are too close to be just a projection effect. 
The large age difference between the binary cluster members point to a formation by tidal capture during a close encounter \citep{van96}. In this sense, Galactic multiple OCs with initial separations smaller than $\sim20$ pc are likely to merge and those with wider separations probably evolve to physically unrelated clusters - both taking place in less than 100 Myr \citep{Portegies07, Fuente10}. 
The fact that the two stellar populations are mixed, even in the central region of each OC and that both parallax and PM distributions are compact and very similar point to an ongoing merger event.
 Although some dispersion is expected for star clusters under tidal disruption, compact stellar distributions in the parallax and PM diagrams are a characteristic of bound structures. The similarity suggests that these objects share the same kinematics.
The Gaia observed and decontaminated stellar surface-density/contour maps (Fig.~\ref{f1}) show clearly  a common structure with two peaks indicating an ongoing merger of a binary cluster.

\begin{table}
{\scriptsize
\begin{center}
\caption{Derived parameters for OCs in the present study.}
\renewcommand{\tabcolsep}{0.7mm}
\renewcommand{\arraystretch}{1.0}
\begin{tabular}{lrrrrr}
\hline
\hline
Cluster&$\alpha(2000)$&$\delta(2000)$&$\sigma_{0K}$&$R_{core}$&$R_{RDP}$\\
&(h\,m\,s)&$(^{\circ}\,^{\prime}\,^{\prime\prime})$&($*\,\arcmin^{-2}$)&($\arcmin$)&($\arcmin$)\\
($1$)&($2$)&($3$)&($4$)&($5$)&($6$)\\
\hline
NGC 1605a &4:34:53.0&45:17:45.0&$5.04\pm0.2$&$1.52\pm0.2$&$10.0\pm3.0$\\
NGC 1605b &4:35:04.0&45:15:57.0&$16.3\pm1.9$&$0.55\pm0.1$&$11.3\pm2.3$\\
Ovd 1 &4:34:10.9&45:25:54.3&$3.97\pm0.1$&$0.75\pm0.1$&$5.0\pm51.0$\\
Ovd 2 &4:34:13.3&45:15:39.5&$8.05\pm1.6$&$0.72\pm0.1$&$6.0\pm2.0$\\
Ovd 3 &4:34:29.0&45:11:35.7&$2.81\pm0.3$&$1.51\pm0.2$&$6.7\pm2.0$\\
Ovd 4 &4:35:26.0&45:02:42.2&$4.82\pm1.3$&$0.72\pm0.2$&$7.0\pm1.5$\\
\hline
\end{tabular}
\label{tab1}
\end{center}
}
\end{table}

During the close encounter, NGC1605a and NGC1605b apparently described looping orbits around each other that are traced by tidal debris deposited along of tidal streams. The Fig.~\ref{f5} shows the streams and tidal debris in an unprecedent way. The binary OC is located within the Perseus arm close to the Galactic disk midplane and, thus, suffer with additional tidal friction from these Galactic features.
Tidal stripping of stars from star clusters has long been expected, including clumps formation \citep{Kupper08}, but the present results suggest that stellar groups can be pulled out from the tidal interacting clusters by drag forces during the close encounter and that some of them may remain gravitationally bound. The structure of these tidal features follows a King profile (Sect.~\ref{sec:2}) and are kinematically coherent.
In addition, the CMDs of the overdensities show stellar populations consistent with those of the two merging clusters suggesting that these stellar debris structures may be fragments of the clusters possibly pulled out by a combination of tidal interaction between the clusters and tidal friction from the Perseus spiral arm and the Galactic disk. Alternatively, evaporated stars could have been coalesced together within the streams forming stellar groups, some of them gravitationally bound. 
There are evidences that the epicyclic motion of escaping stars from star clusters may lead to overdensities formation within the tidal streams, but close encounters with Galactic features like molecular clouds or other star clusters also affect the properties of the tidal debris \citep{Kupper08, Jerabkova21}.
These overdensities could be unrelated OCs, but it is unlikely that they present similar stellar content and follow the binary cluster path by chance. 
 
The Fig.~\ref{f1} shows that the merging clusters have an elongated shape, mostly NGC1605b, which is consistent with tidal tails, but it also can be a new substructure. In the Gaia photometry (right panels) this feature can be seen as a third minor clump. 
Tidal tails have been discovered around globular clusters, but such features around OCs are rare \citep{Minniti18, Roser19, Piatti20, Piatti21}, which probably reflects the difficulty in detecting faint structures within crowding environments.

The RDPs of the binary cluster members and of the four overdensities present a stellar excess in their outskirts. It is primarily due to the presence of the other clusters in the neighbourhood, but,  in part, it also can be interpreted as ongoing dissolution process, in the sense that while the cluster is dissolving stars move out towards the field. Dynamical interactions between star clusters during close encounters increase the internal energy of each cluster at the expense of an orbital kinetic energy reduction. As a consequence of the internal heating by dynamical friction, the clusters' radii increase, while the angular momentum, orbital velocity, and the cluster separation decrease resulting in merger \citep{Priyatikanto16}. There are evidences that in clusters' close encounters the less massive (or dense) component may be completely disrupted, with most of its stars forming a halo around the massive cluster  \citep{Oliveira00}.

Star clusters under tidal stress often have an elongated shape \citep{Chen04}, which may favor their fragmentation during a close encounter. 
It seems to be the case of NGC1605a and b  that even before the mutual encounter probably already had a structure dynamically heated, since they are submitted to the strong Galactic tidal field. 
Due to their shallow gravitational potential OCs are vulnerable against external tidal perturbations and an additional high-energy close encounter may provides enough energy to overcome their binding energy.
In addition, the merger of a cluster pair can result in a fast rotating object \citep{Priyatikanto16}, which increase the mass loss rate \citep{Fuente10, Medina20, Darma21}. 
As aformentioned, Figs. \ref{f1} and \ref{f5} show the tidal debris that trace the orbits followed by the two merging clusters \citep{Leon99, Tang19, Jerabkova21}.
It seems that a previous encounter occurred in the past, probably in the tidal streams crossing shown in the Fig.~\ref{f5} upper and middle panels, $(\Delta{\alpha}',\Delta{\delta})=(-8,5)$. \citet{Priyatikanto16} argue that before a merger takes place, the two OCs typically experience at least one close encounter.

The Fig.~\ref{f5}, bottom panel, shows the overdensities 1 and 2. The tidal debris distribution around these two merger relics are consistent with the disruption of rotating structures  \citep{Penarrubia09, Jerabkova21}.
There are secondary streams around these overdensities that suggest looping orbits of tidal substructures possibly stripped from them.
It's easy to accept that an elongated and fast rotating OC tends to lose stars preferentially at its ends with the stripped stars drawing looping orbits around the cluster. For a high-energy close encounter the possibility of fragmentation of such elongated OCs, already weakened by the Galactic tidal effects, cannot be ruled out.  A future simulation could provides clues on the merger process that leaded up to the complex scenario shown in Fig.~\ref{f5}.

A stellar bridge connecting NGC1605a and NGC1605b (Fig.~\ref{f1}- middle panel) suggests a possible mass transfer \citep{Dieball98}. This feature may explain the significant fraction of stars from NGC1605b in the FS decontaminated CMD of the central region of NGC1605a, in addition to possible previous encounters. 

It is commonly accepted that merging events lead to the formation of massive clusters, however, it is interesting to speculate that such processes, at least when they result from close encounters, can break down massive OCs into multiple low-mass objects by tidal stripping, some of them eventually result bound. These features are potentially interesting for future simulations, as they can be completely dissolved in the Galactic field due to tidal heating \citep{Pang21}, also may result in a cluster aggregate/moving group \citep{Dalessandro21}, or disperse as bound structures following their own orbits in the Galaxy. Recently, \citet{Coronado21} identified several OCs sharing the same orbit.

\section{Concluding remarks}
\label{sec:5}

This paper reports the identification of an old binary OC pair in the Milky Way. The binary cluster members are NGC1605a with $2\pm0.2$ Gyr and NGC1605b with $600\pm100$ Myr, both located at the same distance from the Sun of $d_{\odot}=2.6\pm0.4$ kpc and with a projected separation of only $\sim1.8$ pc between their central cores.
These parameters suggest a physical connection and the age difference point to a formation by tidal capture during a close encounter.  
The 2MASS and Gaia-EDR3 field-stars decontaminated CMDs of these objects show two distint populations that are mixed within a large area around the binary cluster including the central region of each one, which suggest an ongoing merger event. 
The compact and very similar parallax and PM distributions also are consistent with an ongoing merger of a binary cluster pair.

The merging clusters are followed by tidal streams that appear to draw their recent orbital path. These tidal streams are populated by several stellar overdensities, four of them result in OCs - Camargo 1010, 1011, 1012, and 1013.
These tidal structures are populated by stars consistent with those in the merging clusters with the adopted isochrone solutions for NGC1605a and NGC1605b fitting very well the two stellar populations within these overdensities. These new findings are distributed along of the stream that seems to be the NGC1605b orbit and, thus, the age of $600$ Myr was adopted for them. The heliocentric distance derived for these newly found OCs is the same of the binary cluster. Also, these overdensities share the same PM distribution as the merging clusters. 

This study suggests that during a close encounter of two OCs, in addition to the evaporation of individual stars from these clusters, stellar groups  can be pulled out of them by tidal effects, forming tidal debris distributed along of tidal streams and, surprisingly, some of them may survive as bound stellar systems. 

From the theoretical viewpoint, this binary cluster may be an important seed on developing and validate more realistic simulations for tidal interacting stellar systems. 

Close encounters between star clusters are rare, obviously the subsequent formation of binary clusters is even rarer, and the evolution to a merger event is extremely unlikely, which makes the present discovery extremely important.
NGC1605a and NGC1605b become the first old binary OC known in the Galaxy and, if that was not enough, they appear to be undergoing a merger during a close encounter, leaving streams populated by bound substructures. In addition, the new finding alleviates the scarcity of binary cluster pairs in our home Galaxy when compared to the Magellanic Clouds.

\vspace{0.8cm}

\textit{Acknowledgements}: The author thank the anonymous referee and the editor for useful comments and suggestions that greatly improve the quality of the manuscript. 
This publication makes use of data products from WISE, 2MASS, and Gaia-EDR3. 
The Wide-field Infrared Survey Explorer is a joint project of the University of California, Los Angeles and JPL/CIT funded by NASA.
The Two Micron All Sky Survey is a joint project of the University of Massachusetts and the IPAC/CIT, funded by NASA and the National Science Foundation. 
This work also has made use of data from the European Space Agency (ESA) mission Gaia (\url{https://www.cosmos.esa.int/gaia}), processed by the Gaia Data Processing and Analysis Consortium 
(\url{DPAC, https://www.cosmos.esa.int/web/gaia/dpac/consortium}). Funding for the DPAC has been provided by national institutions, in particular the institutions participating in the Gaia Multilateral Agreement.

\end{document}